\documentclass[12pt]{article}

\begin{document}
\input amssym.tex

\title{How to kill the Unruh effect}
\author{Ion I.  Cot\u aescu\thanks{E-mail:~~~cota@physics.uvt.ro}  \\
{\small \it West University of Timi\c soara,}\\
{\small \it V.  P\^ arvan Ave.  4, RO-300223 Timi\c soara, Romania}}

\maketitle

\begin{abstract}
How to kill the Unruh effect? Very simple, by requiring the Rindler transformation to behave continuously for vanishing acceleration. Then the Unruh effect disappears as we show in the case of the massive scalar quantum field. The main point is that the continuity condition restricts the integral of the mode expansion to the positive energy spectrum, suppressing thus the $\beta$ - terms of the Bogoliubov transformation.  

Pacs{04.62.+v}
\end{abstract}

\maketitle


\newpage

\section{Introduction}

The study of the Unruh effect \cite{unruh} is one of the most popular fields of research being considered by many authors as a  paradigm opening new horizons in investigating cosmological problems \cite{crispino}. However, in our opinion this effect is rather the consequence of a too broad freedom and arbitrariness in defining quantum modes without taking care on the minimal restrictions required by the coherence of the quantum theory.  

The Unruh problem addresses the experience of two moving observers, among them the first one (${\frak M}$) stays in an inertial frame of the Minkowski spacetime, while the second one (${\frak R}$) moves with an uniform accelerated non-inertial frame complying with a Rindler transformation, i.e., the Rindler wedge \cite{BD}. Each observer has his own apparatus which can prepare and measure quantum states in his own frame, where the apparatus is represented by the entire algebra of observables defined in this frame. The question is how can an observer measure the quantum states prepared by the other observer using his own apparatus. In other words, how the accelerated observer ${\frak R}$ measures the quantum modes prepared in the inertial frame by  ${\frak M}$.    

This catching problem was intensively studied during the last decades considering either the mentioned intrinsic apparatus or special Unruh-DeWitt detectors \cite{unruh_dw}.  Nevertheless, many results, controversial or not \cite{fedotov}, were obtained using indirect arguments \cite{takagi} instead of  mathematical demonstrations. The first rigorous derivation of the Bogoliubov coefficients between the Minkowskian modes of the inertial frame and the Fulling ones \cite{fulling} defined on the Rindler wedge was performed recently by Longhi and Soldati \cite{longi_soldati}. Obviously, they confirm the previous results concerning the same combination of modes, where it seems that the presence of the Fulling ones is crucial because their good asymptotic properties. However, we may ask if this is the only possible choice we have for assuring the global convergence of the theory. 

The problem is not trivial since there are no explicit principles which should guide us in interpreting the measurements of two completely independent and unrestricted observers. However, in the case of the isometry transformations, where we also have  at least two frames with specific observers, this problem is implicitly solved by using Lie groups. In this case, the transformations depend continuously on parameters which vanish when  the isometry reduces to the identity one. Physically speaking this is in fact a fine tuning between both the observer's apparatus which have to record the same data when the isometry parameters vanish.

Why is this elementary condition of continuity  ignored in the case of the  Unruh problem? Because of technical difficulties or due to some conjunctures, when does this limit not make sense? Anyway, we have now the tools allowing us to face this challenge \cite{longi_soldati}. In this paper we would like to study the consequences of the continuity of the Unruh problem for vanishing acceleration. More specific, we assume that any quantity measured by ${\frak R}$ at acceleration $a$, denoted by $q_R(a)$, must have as a limit for $a=0$ precisely the equivalent quantity $q_M$ measured by ${\frak M}$, i.e., 
\begin{equation}
\lim_{a\to 0} q_R(a)=q_M\,, 
\end{equation}
for any $q$, including the quantum modes. In order to work out this problem we start with continuous Rindler transformations which tend to the identity one when   $a\to 0$. Moreover, we require the mode functions measured by  ${\frak R}$ at $a=0$ to coincide with those prepared by  ${\frak M}$.

We must stress that the Fulling modes have no limits for $a=0$ and, therefore, we need to introduce new modes accomplishing the continuity condition. Our principal result is that in the Rindler wedge the  mode expansion of the scalar field must be written  integrating only over the positive energy spectrum, preserving thus the vacuum stability.  However, the price to pay for the good continuity properties is to involve strongly divergent mode functions. The problem is if these divergences could be somehow eliminated at least from the brackets giving physical results. This problem is too complicated to be solved here  even though we suggest some possibilities of doing this.

\section{Minkowskian modes}

Let us consider the massive and charged scalar quantum field, $\phi:M\to {\Bbb C}$, defined on the Minkowski spacetime, $M$,  and satisfying the Klein-Gordon equation, $(\square +m^2)\phi=0$. In the inertial frame $\{t,x\}$ of the observer ${\frak M}$, this field can be expanded in  momentum representation,
\begin{equation}\label{field}
\phi(t,x)=\phi_{+}(t,x)+\phi_{-}(t,x)=\int_{-\infty}^{\infty} dp \left[f_p(t,x)a(p)+{f_p(t,x)}^*a^c(p)^{\dagger}\right]
\end{equation}
in terms of field operators, $a$ and $a^c$, and mode functions,
\begin{equation}
f_p(t,x)=\frac{1}{\sqrt{2\pi}}\frac{e^{-iEt+ipx}}{\sqrt{2E}}\,, \quad E=\sqrt{p^2+m^2}\,,
\end{equation}
that are orthonormal with respect to the Minkowskian relativistic scalar product
\begin{equation}\label{SP1}
\langle \phi,\phi'\rangle_M=i\int_{\Sigma} d\sigma^{\mu}\,\phi^*
\stackrel{\leftrightarrow}{\partial_{\mu}} \phi'=i\int_{-\infty}^{\infty} dx\,\phi^*
\stackrel{\leftrightarrow}{\partial_{t}} \phi'\,,
\end{equation}
calculated on $\Sigma={\Bbb R}$ (at $t={\rm const}.$) using the notation $f\stackrel{\leftrightarrow}{\partial}h=f(\partial h)-h(\partial f)$.  Since the mode functions are orthonormalized in the momentum scale,
\begin{eqnarray}
\langle  f_{p},f_{{p}'}\rangle_M=-\langle  f_{p}^*,f_{{
p}'}^*\rangle_M&=&\delta({p}-{p}')\,,\\
\langle f_{p},f_{{p}'}^*\rangle_M&=&0\,,\label{fstf}
\end{eqnarray}
the field $\phi$ is canonically quantized if the field operators obey the non-vanishing commutation rules
\begin{equation}
\left[a(p),{a(p')}^{\dagger}\right]=\delta(p-p')\,,\quad 
\left[a^c(p),{a^c(p')}^{\dagger}\right]=\delta(p-p')\,.
\end{equation} 

We remind the reader that the separation between the positive and negative frequencies splits the Hilbert space of the square integrable mode functions, 
${\cal H}={\cal H}_+\oplus {\cal H}_-$.  The set of fundamental solutions 
$\{ f_p\, |\,p\in {\Bbb R}\}$ forms a (generalized) basis in ${\cal H}_+$ while their complex conjugate functions,  $\{ f_p^*\, |\,p\in {\Bbb R}\}$, represent a basis in ${\cal H}_-\equiv {\cal H}_+^*$. Obviously, Eq. (\ref{fstf}) guarantees the orthogonality of these subspaces.

In investigating the Unruh problem it is interesting to analyse separately what happens with the progressive and regressive waves \cite{crispino}. For this reason we use the Heaviside step function $\theta$ for defining the progressive $(+)$ and regressive $(-)$ wave functions and the corresponding field operators,
\begin{equation}
f^{(\pm)}_p=\theta(p)f_{\pm p}\,,\quad a^{(\pm)}(p)=\theta(p)a(\pm p)\,,\quad
a^{c\,(\pm)}(p)=\theta(p)a^c(\pm p)\,, 
\end{equation}
which allow us to write 
\begin{equation}
\phi(t,x)=\int_{0}^{\infty} dp \left[f^{(+)}_p(t,x)a^{(+)}(p)+{f^{(-)}_p}t,(x)a^{(-)}(p)\right] +\,{\rm neg. freq.}\,. 
\end{equation}
We observe that the new wave functions are orthogonal since these satisfy 
\begin{equation}
\langle  f_{p}^{(\pm)},f^{(\pm)}_{p'}\rangle_M=\delta(p-p')\,,\quad
\langle  f_{p}^{(\pm)},f^{(\mp)}_{p'}\rangle_M=0\,,\quad {\rm etc.}
\end{equation}
Thus we split the subspaces ${\cal H}_\pm={\cal H}^{(+)}_\pm\oplus {\cal H}^{(-)}_\pm$ laying out the orthogonal subspaces of progressive  and regressive wave functions. Then we have 
\begin{equation}
\left[a^{(\pm)}(p),{a^{(\pm)}(p')}^{\dagger}\right]=\theta(p)\delta(p-p')\,,\quad 
\left[a^{(\pm)}(p),{a^{(\mp)}(p')}^{\dagger}\right]=0\,,
\end{equation} 
and similar for $a^{c\,(\pm)}(p)$.

The quantum field can also be expanded using the wave functions normalized in the energy scale,
\begin{equation}\label{Mmodes}
f^{(\pm)}_E(t,x)=\sqrt{\frac{E}{p}}f^{(\pm)}_p(t,x)=\frac{1}{\sqrt{2\pi}}\frac{e^{-iEt\pm ipx}}{\sqrt{2p}}\,,\quad \forall E\ge 0\,,
\end{equation}
which satisfy now
\begin{equation}
\langle  f_{E}^{(\pm)},f^{(\pm)}_{E'}\rangle_M=\delta(E-E')\,,\quad
\langle  f_{E}^{(\pm)},f^{(\mp)}_{E'}\rangle_M=0\,,\quad {\rm etc.}
\end{equation}
since $\delta(E+E')=0$ when $E,E'> 0$. The field operators normalized in the energy scale, 
\begin{equation}
a^{(\pm)}(E)=\sqrt{\frac{E}{p}}\,a^{(\pm)}(p )\,,\quad a^{c\,(\pm)}(E)=\sqrt{\frac{E}{p}}\,a^{c\,(\pm)}(p )\,,
\end{equation}
accomplish,
\begin{equation}
\left[a^{(\pm)}(E),{a^{(\pm)}(E')}^{\dagger}\right]=\delta(E-E')\,,\quad 
\left[a^{(\pm)}(E),{a^{(\mp)}(E')}^{\dagger}\right]=0\,.
\end{equation} 
and similar for $a^{c\,(\pm)}(E)$.

Now we adopt the position of the observer ${\frak R}$ denoting the all quantities defined above by the observer ${\frak  M}$ with the index $_M$. Thus in the previous formulas we replace  $\{t,x\} \to\{t_M,x_M\}$, and denote the energy and momentum operators by $H_M=i\partial_{t_M}$ and $P_M=-i\partial_{x_M}$ respectively, bearing in mind that their corresponding eigenvalues,  $\tilde E$ and $\tilde p$, satisfy $\tilde E^2=\tilde p^2+m^2$. With these new notations, we obtain the definitive expansion in the energy scale,
\begin{eqnarray}
\phi(t_M,x_M)&=&\int_{m}^{\infty} d\tilde E \left[f^{(+)}_{\tilde E}(t_M,x_M)\,a^{(+)}(E)+{f^{(-)}_{\tilde E}}(t_M,x_M)\,a^{(-)}(\tilde E)\right. \nonumber\\
&+&\left. f^{(+)\,*}_{\tilde E}(t_M,x_M)\,{a^{c\,(+)}(\tilde E)}^{\dagger}+{f^{(-)\,*}_{\tilde E}}(t_M,x_M)\,{a^{c\,(-)}(\tilde E)}^{\dagger}\right]\,,\label{Mexp}
\end{eqnarray}
integrating  only over the positive energy spectrum $[m,\infty)$.

\section{Continuous Rindler transformations}

We assume that the observer ${\frak R}$ stays in the (right) Rindler wedge, ${\cal R}(a)$, of coordinates $\{t,x\}$ defined in an unusual manner as, 
\begin{equation}\label{coord}
t_M=\frac{1}{a}\,e^{ax}\sinh at\,, \quad x_M=\frac{1}{a}\left(e^{ax}\cosh at-1\right)\,,
\end{equation}
since the continuity condition in $a=0$ requires,  
\begin{equation}
\lim_{a\to 0} t_M=t\,, \quad \lim_{a\to 0} x_M=x\,.
\end{equation}
We say that this is the {\em continuous} Rindler transformation. The inverse transformation is now of the form
\begin{equation}
t=\frac{1}{a}\,{\rm arctanh}\frac{a t_M}{a x_M+1}\,,\quad x=\frac{1}{2a}\, \ln\left[\left( a x_M+1\right)^2-(a t_M)^2\right]\,,
\end{equation}
that holds since $x_M+\frac{1}{a}>|t_M|$. We specify that here we do not speak about the left Rindler wedge since this does not play any role as long as $\lim_{a\to 0}{\cal R}(a)=M$.  

The extra term of Eq. (\ref{coord}b) represents a translation that does not affect the line element,
\begin{equation}
ds^2=dt_M^2-dx_M^2=e^{2ax}(dt^2-dx^2)\,.
\end{equation}
and the Klein-Gordon equation, 
\begin{equation}\label{KGR}
\left(\partial_t^2-\partial_x^2+m^2 e^{2ax}\right)\phi(t,x)=0\,,
\end{equation} 
which allows the observer ${\frak R}$ to derive the quantum modes on ${\cal R}(a)$. 

The principal operators are now the Rindler energy,  
\begin{equation}
H = i\partial_t=H_M+a(x_M H_M-t_M P_M)\,,
\end{equation} 
and momentum,
\begin{equation}
P =-i\partial_x=P_M+a(x_M P_M-t_M H_M)\,,
\end{equation}
which satisfy the continuity condition when $a\to 0$ (i.e., $H\to H_M$ and $P\to P_M$). We observe that the operator $H$ is conserved commuting with $\square$ since its second term is just the generator of the Lorentz transformations along the $x$-axis. Thus the observer ${\frak R}$  can use the set of commuting operators $\{\square,H\}$ for determining quantum modes. Notice that these are no longer eigenfunctions of $P$ since this operator does not commute with $\square$.

The scalar mode functions must be orthogonal with respect to the Rindler scalar product 
\begin{equation}\label{SP2}
\langle \phi,\phi'\rangle_R=i\int_{\Sigma'} d\sigma^{\mu}\,\phi^*
\stackrel{\leftrightarrow}{\partial_{\mu}} \phi'=i\int_{-\infty}^{\infty} dx\,\phi^*\stackrel{\leftrightarrow}{\partial_{t}} \phi'\,,
\end{equation}
calculated on $\Sigma'\not=\Sigma$ defined now by $t={\rm const}.$ (instead of $t_M={\rm const}.$). Thus each observer has its own scalar product following we see how these are related among themselves when the Rindler transformations are continuous. To this end we calculate first
\begin{equation}
\langle f_{\tilde E}^{(\pm)},f_{{\tilde E}'}^{(\pm)}\rangle_R =i\int_{-\infty}^{\infty} dx {f^{(\pm)}_{\tilde E}[t_M(t,x),x_M(t,x)]}^*\stackrel{\leftrightarrow}{\partial_{t}}   f^{(\pm)}_{\tilde E'}[t_M(t,x),x_M(t,x)]
\end{equation} 
taking the integral at $t=0$ since this is anyway time-independent \cite{BD}. Changing then the variable we obtain the integrals
\begin{equation}\label{scalar2}
\langle f_{\tilde E}^{(\pm)},f_{{\tilde E}'}^{(\pm)}\rangle_R=\frac{e^{\mp\frac{i}{a}(\tilde p'-\tilde p)}}{4\pi \sqrt{\tilde p\tilde p'}}\int _{0}^{\infty}d\xi\, e^{\pm i(\tilde p'-\tilde p)\,\xi}\,, \quad \xi=\frac{1}{a}\,a^{ax}\,,
\end{equation}
that can be combined giving the following identities:
\begin{eqnarray}
\langle f_{\tilde E}^{(\pm)},f_{{\tilde E}'}^{(\pm)}\rangle_M&=&e^{\pm\frac{i}{a}(\tilde p'-\tilde p)}\langle f_{\tilde E}^{(\pm)},f_{\tilde E'}^{(\pm)}\rangle_R+e^{\mp\frac{i}{a}(\tilde p'-\tilde p)}\langle f_{\tilde E}^{(\mp)},f_{{\tilde E}'}^{(\mp)}\rangle_R\,,\label{identSC1}\\
\langle f_{\tilde E}^{(\mp)},f_{{\tilde E}'}^{(\pm)}\rangle_M&=&e^{\pm\frac{i}{a}(\tilde p'+\tilde p)}\langle f_{\tilde E}^{(\mp)},f_{\tilde E'}^{(\pm)}\rangle_R+e^{\mp\frac{i}{a}(\tilde p'+\tilde p)}\langle f_{\tilde E}^{(\pm)},f_{{\tilde E}'}^{(\mp)}\rangle_R\,.\label{identSC2}
\end{eqnarray} 
Notice that the first one permits us to calculate normalization factors using the rule $\langle\,\, ,\,\rangle_R=\frac{1}{2}\langle\,\, ,\,\rangle_M$ that holds since the Minkowskian modes are normalized alike at $\tilde p' =\tilde p$. The second identity is merely formal as long as all its brackets vanishes as $\delta(p+\tilde p)$ with $p,\tilde p >0$.   

\section{Rindler modes}

The Rindler modes are linear combinations of fundamental solutions of Eq. (\ref{KGR}) which are simultaneously eigenfunctions of the energy operator $H$, whose eigenvalues are denoted by $E$. The solutions of positive frequencies may be any linear combination of the divergent functions
\begin{equation}
I_{\pm i\frac{E}{a}}\left(\frac{m}{a}\,e^{ax}\right)e^{-iEt}\,, 
\end{equation}
where $I$ are the modified Bessel functions presented in Appendix. We remind the reader that the Fulling modes \cite{fulling}
\begin{equation}
K_{i\frac{E}{a}}\left(\frac{m}{a}\,e^{ax}\right)e^{-iEt}\,,
\end{equation}
are the only linear combinations (\ref{modbes}b) with a good asymptotic behaviour, preferred in many studies concerning the Unruh effect \cite{crispino}. However, as mentioned, we try here to find the mode functions that become  the Minkowskian ones in the limit of $a\to 0$. The problem is not trivial since the asymptotic behaviour of the modified Bessel functions is complicated as we can see from Eqs. (\ref{lim1}) and (\ref{lim2}) that yield 
\begin{eqnarray}
I_{\pm i\frac{E}{a}}\left(\frac{m}{a}\,e^{ax}\right)&=&\frac{\sqrt{a}}{\sqrt{2\pi}}\frac{e^{\frac{\pi E}{2a}}}{(E^2-m^2 e^{2ax})^{\frac{1}{4}}}\,e^{\pm i\vartheta(a,x)}+{\cal O}(a)\,,\\ 
K_{i\frac{E}{a}} \left(\frac{m}{a}\,e^{ax}\right)&=&-\sqrt{2\pi a\,}\frac{e^{-\frac{\pi E}{2a}}}{(E^2-m^2 e^{2ax})^{\frac{1}{4}}}\sin \vartheta(a,x)+{\cal O}(a)\,.
\end{eqnarray}
The phase function
\begin{equation}\label{phaseax}
\vartheta(a,x)=\frac{1}{a}\sqrt{E^2-m^2 e^{2ax}}-\frac{E}{a}\, {\rm arccosh}\, \frac{E e^{-ax}}{m}-\frac{\pi}{4}
\end{equation} 
allows the series expansion 
\begin{equation}
\vartheta(a,x)=\frac{p}{a} -\frac{E}{a}\,{\rm arcsinh}\,\frac{p}{m} + 
p\, x - \frac{\pi}{4} + {\cal O}(a)\,,\quad  p=\sqrt{E^2-m^2}\,,
\end{equation}
showing that this has a pole in $a=0$. Consequently, correct limits for $a\to 0$ can be obtained only when the divergence generated by this pole is removed. We must stress that this cannot be done in the case of the Fulling modes since  the $\sin$ function involving the divergent phase function (\ref{phaseax}) does not have a limit for $a\to 0$.

The solution is to define {\em separately} the progressive $(+)$ and regressive $(-)$ Rindler mode functions of positive frequencies as   
\begin{equation}\label{Rmodes}
R^{(\pm)}_{E}(t,x)=N\,I_{\pm i\frac{E}{a}}\left(\frac{m}{a}\,e^{ax}\right)e^{\mp i\vartheta_0}e^{-iEt}\,,\quad \forall\, E\ge 0\,, 
\end{equation}
where the fixed phase
\begin{equation}
\vartheta_0=\frac{p}{a} -\frac{E}{a}\,{\rm arcsinh\,}\frac{p}{m}-\frac{\pi}{4}\,,  
\end{equation}
removes the singularities at $a=0$. We note that these mode functions are defined only for $E\in [m, \infty)$ since otherwise the phase factor becomes a real factor that changes the norm. Furthermore, observing that the equality  $\langle\,\, ,\,\rangle_R=\frac{1}{2}\langle\,\, ,\,\rangle_M$ holds in this case,  we impose the prescription 
\begin{equation}\label{ortoR1}
\lim_{a\to 0}\langle R^{(\pm)}_E, R^{(\pm)}_{E'}\rangle_R=\frac{1}{2}\,\delta(E-E')\,,\quad 
\lim_{a\to 0}\langle R^{(\pm)}_E, R^{(\mp)}_{E'}\rangle_R=0\,,
\end{equation}
determining  the normalization constant
\begin{equation}\label{N}
N=\frac{1}{2\sqrt{a \sinh\frac{\pi E}{a}}}\,.
\end{equation}
Finally, we verify the desired continuity condition,
\begin{equation}
\lim_{a\to 0} R^{(\pm)}_E(t,x)=f_E^{(\pm)}(t,x)\,,
\end{equation}
which indicates that for small values of $a$  we have $R^{(\pm)}_E\in {\cal H}_+$ and $R^{(\pm)\,*}_E\in {\cal H}_-$. 

The scalar field in Rindler wedge, $\phi(t,x)=\phi[t_M(t,x),x_M(t,x)]$, can be expanded now in terms of new field operators, $b^{(\pm)}(E)$ and $b^{c\,(\pm)}(E)$, as  
\begin{eqnarray}\label{scalarR}
\phi(t,x)&=&\int_{m}^{\infty}dE\left[R^{(+)}_E(t,x)\,b^{(+)}(E)+R^{(-)}_E(t,x)\,b^{(-)}(E)\right.\nonumber\\
&&~~~ \left. +\, R^{(+)\,*}_E(t,x)\,{b^{c\,(+)}(E)}^{\dagger}+R^{(-)\,*}_E(t,x)\,{b^{c\,(-)}(E)}^{\dagger}\right]\,.\label{Rexp} 
\end{eqnarray}
We specify that continuity at $a=0$ forces us to integrate only over the positive energy spectrum.  Therefore, we expect the continuous Rindler transformations to do not mix up the states of positive and negative frequencies, preserving thus the same vacuum state in both the charts under consideration.   

The prescription (\ref{ortoR1}) requires to have  
\begin{equation}\label{limitbb}
\lim_{a\to 0}\left[b^{(\pm)}(E),{b^{(\pm)}(E')}^{\dagger}\right]=\delta(E-E')\,,\quad 
\lim_{a\to 0}\left[b^{(\pm)}(E),{b^{(\mp)}(E')}^{\dagger}\right]=0\,.
\end{equation} 
and similar for $b^{c\,(\pm)}(E)$. On the other hand, the expansion (\ref{scalarR}) involves strongly divergent mode functions which can compromise the physical meaning. The problem is to find suitable commutation relations of the field operators which able to assure the canonical commutation rules of the scalar field. O course, we can not give an answer here but we observe that there are many interesting possibilities compatible with the conditions (\ref{limitbb}). For example, we can use commutation relations of the form,
\begin{eqnarray}
\left[b^{(\pm)}(E),{b^{(\pm)}(E')}^{\dagger}\right]&=&\frac{\kappa}{a\sqrt{\pi}}\,e^{-\frac{\kappa2}{a^2}(E-E')^2}+{\cal O}(a)\,,\\\label{comutbb} 
\left[b^{(\pm)}(E),{b^{(\mp)}(E')}^{\dagger}\right]&=&\frac{\kappa}{a\sqrt{\pi}\,}e^{-\frac{\kappa2}{a^2}(E+E')^2}+{\cal O}(a)\,,\label{comutbb1}
\end{eqnarray}  
where $\kappa$ is an arbitrary constant. It remains to see if (and how)  such rules could eliminate the effects due to the divergent Rindler modes considered here.

\section{Transition coefficients}

The Rindler field operators $b^{(\pm)}$ and $b^{c\,(\pm)}$ are related to the Minkowskian ones,  $a^{(\pm)}$ and $a^{c\,(\pm)}$, through the transition coefficients that transform the Minkowskian and Rindler bases, i.e., $\langle f^{(+)}_{\tilde E}, R^{(+)}_E\rangle_M$, $\langle f^{(-)}_{\tilde E}, R^{(+)}_E\rangle_M$ ...etc.. These may be derived starting with the Rindler ones, $\langle f^{(+)}_{\tilde E}, R^{(+)}_E\rangle_R$, $\langle f^{(-)}_{\tilde E}, R^{(+)}_E\rangle_R$ ..., and assuming that identities of the form (\ref{identSC1}) and (\ref{identSC2}) hold even in this case. For example, the first bracket, 
\begin{equation}
\langle f^{(+)}_{\tilde E}, R^{(+)}_E\rangle_R = i\int_{-\infty}^{\infty} dx {f^{(+)}_{\tilde E}[t_M(t,x),x_M(t,x)]}^*\stackrel{\leftrightarrow}{\partial_{t}}R_E^{(+)}(t,x)\,, 
\end{equation}   
can be calculated at $t=0$ as in Ref. \cite{longi_soldati}. According to Eqs. (\ref{Mmodes}), (\ref{Rmodes}) and (\ref{N}), we have 
\begin{eqnarray}
\langle f^{(+)}_{\tilde E}, R^{(+)}_E\rangle_R&=& \frac{e^{-i\vartheta_0+i\frac{\tilde p}{a}}}{4\sqrt{\pi a \tilde p \sinh\frac{\pi E}{a}}}\nonumber\\
&&\times \int_{-\infty}^{\infty} dx \left[E+\tilde E e^{ax}\right]e^{-i \frac{\tilde p}{a}e^{ax}} I_{i\frac{E}{a}}\left(\frac{m}{a}\,e^{ax}\right)\,, 
\end{eqnarray}
where $\tilde p=\sqrt{{\tilde E}^2-m^2}$. The last step is to change  the variable $x$ into $\xi$ as given by Eq. (\ref{scalar2}b). Then the above integral  takes the form,
\begin{equation}
\int_{0}^{\infty} d\xi \left[\frac{E}{a}\frac{1}{\xi}+\tilde E \right]e^{-i {\tilde p}\xi} I_{i\frac{E}{a}}\left(m\xi\right)\,, 
\end{equation}
involving integrals as (\ref{intA}) and (\ref{intB}). Unfortunately, the parameters of these integrals are far from the domain of convergence. Nevertheless, for vanishing mass we may replace $\tilde p\to \tilde p - i \epsilon$, $E \to E-i \epsilon$ and $m\to \frac{\epsilon}{2}$ restoring the convergence.  Then, for small values of $\epsilon$, we estimate  
\begin{eqnarray}
\langle f^{(\pm)}_{\tilde E}, R^{(+)}_E\rangle_R&\sim& \frac{e^{\frac{\pi E}{2a}}}{2\sqrt{\pi a \tilde p \sinh\frac{\pi E}{a}}}\,,\\
\langle f^{(\pm)}_{\tilde E}, R^{(-)}_E\rangle_R&\sim& \frac{e^{-\frac{\pi E}{2a}}}{2\sqrt{\pi a \tilde p \sinh\frac{\pi E}{a}}}\,,\\
\langle f^{(\pm)\,*}_{\tilde E}, R^{(+)}_E\rangle_R&=&\langle f^{(\pm)}_{\tilde E}, R^{(+)\,*}_E\rangle_R=0\,,\\
\langle f^{(\pm)\,*}_{\tilde E}, R^{(-)}_E\rangle_R&=&\langle f^{(\pm)}_{\tilde E}, R^{(-)\,*}_E\rangle_R=0\,.
\end{eqnarray}
This confirms that the continuous Rindler transformations mix the progressive and regressive modes among themselves but without affecting the vacuum stability.

\section{Concluding remarks}

The principal conclusion is that our continuity prescription guarantees the vacuum stability eliminating the traditional Unruh effect generating thermal baths. The Unruh conjecture becomes now more realistic involving two observers, an inertial and an accelerated one, which measure different beam intensities observing the same particle beam prepared by one of them, but without perceiving antiparticles in the same time.

Why is a simple continuity condition able to suppress the $\beta$ -terms of the Bogoliubov transformations?  In our opinion this is because its use disciplines  the calculation procedure, preventing us to integrate abusively over the negative energies as it commonly happens \cite{crispino,longi_soldati}.

Finally, we note that our approach opens the perspective of predicting new  physical effects in accelerated frames. The principal difficulty here comes from the divergent functions $I_{\pm i\nu}$ generating singular terms that could be removed using new types of commutation rules as in (\ref{comutbb}) and (\ref{comutbb1}). Otherwise, it remains to use only the Fulling modes but respecting the integration on the positive energies which eliminates the Unrruh effect.

\subsection*{Appendix: Modified Bessel functions}

\setcounter{equation}{0} \renewcommand{\theequation}
{A.\arabic{equation}}

Let us consider the modified Bessel functions of pure imaginary indices
$I_{i\nu}$ and $K_{i\nu}$  defined as
\begin{equation}\label{IJK}
I_{i\nu}(x)=e^{\frac{1}{2}\pi\nu}J_{i\nu}(ix)\,,\quad K_{i\nu}(x)=i\frac{\pi}{2}
\frac{I_{i\nu}(x)-I_{-i\nu}(x)}{\sinh\pi\nu}\,,~~x,\,\nu\in {\Bbb R}\,,
\end{equation}
which have the remarkable properties $(I_{i\nu})^*=I_{-i\nu}$ and $(K_{i\nu})^*=K_{-i\nu}=K_{i\nu}$. For large values of $x,\nu>0$ these functions behave as \cite{bateman}
\begin{eqnarray}
I_{\pm i\nu}(x)&=&\frac{1}{\sqrt{2\pi}}\frac{e^{\frac{1}{2}\pi\nu}}{(\nu^2-x^2)^{\frac{1}{4}}}\,e^{\pm i\vartheta_{\nu}(x)}+{\cal O}(x^{-1})\,,\label{lim1}\\
K_{i\nu}(x)&=&-\sqrt{2\pi\,}\frac{e^{-\frac{1}{2}\pi\nu}}{(\nu^2-x^2)^{\frac{1}{4}}}\sin \vartheta_{\nu}(x)+{\cal O}(x^{-1})\,,\label{lim2}
\end{eqnarray}
where
\begin{equation}\label{phase}
\vartheta_{\nu}(x)=\sqrt{\nu^2-x^2}-\nu\, {\rm arccosh} \frac{\nu}{x}-\frac{\pi}{4}
\end{equation}

There are two related integrals which can be expressed in terms of elementary functions \cite{gradstein},
\begin{eqnarray}\label{modbes}
{\cal J}&\equiv&\int_0^{\infty}\frac{dx}{x}\,e^{-i\alpha x}I_{i\nu}(\beta x)= \frac{-i\,e^{\frac{1}{2}\pi\nu}}{\nu}\left(\frac{1}{\beta}\sqrt{\alpha^2+\beta^2}-\frac{\alpha}{\beta}\right)^{i\nu}\,, \label{intA}\\
i\partial_{\alpha}{\cal J}&\equiv&\int_0^{\infty}dx\,e^{-i\alpha x} I_{i\nu}(\beta x)= \frac{-i\,e^{\frac{1}{2}\pi\nu}}{\sqrt{\alpha^2+\beta^2}}\,\left(\frac{1}{\beta}\sqrt{\alpha^2+\beta^2}-\frac{\alpha}{\beta}\right)^{i\nu}\,.\label{intB}
\end{eqnarray}
These integrals converge for $\Im \nu<0$ and $\Im \alpha <-\Re \beta$.

\end{document}